\RequirePackage[compress]{cite} 
\documentclass[final,3p,linenumbers,compress]{elsarticle}

\makeatletter
\newcommand*{\rom}[1]{\expandafter\@slowromancap\romannumeral #1@}
\makeatother

\usepackage{lineno,hyperref}
\usepackage{multirow}

\usepackage{amssymb}
\usepackage{amsmath}
\usepackage{graphicx,color}

\usepackage{cite}

\usepackage{multicol}

\usepackage{float}

\usepackage{xspace}	
\usepackage{caption}
\usepackage{float}

\journal{Physics Letters B}

\DeclareUnicodeCharacter{2212}{-}

\newcommand{\roots} {\mbox{$\sqrt{\textit{s}_{NN}}$}}

\def  \vn         {\mbox{$\textit{v}_{n}$}} 
\def  \first      {\mbox{$\textit{v}_{1}$}}
\def  \second     {\mbox{$\textit{v}_{2}$}}
\def  \third      {\mbox{$\textit{v}_{3}$}}

\def  \etas       {\mbox{$\eta / \textit{s}$  }}

\def  \nth       {\mbox{\textit{n}$^{\textrm{th}}$}}
\begin{document}

\begin{frontmatter}

\title{Beam-energy dependence of correlations between mean transverse momentum and anisotropic flow of charged particles in Au+Au collisions at RHIC}
\author{
B.~E.~Aboona$^{59}$,
J.~Adam$^{17}$,
G.~Agakishiev$^{32}$,
I.~Aggarwal$^{45}$,
M.~M.~Aggarwal$^{45}$,
Z.~Ahammed$^{66}$,
A.~Aitbayev$^{32}$,
I.~Alekseev$^{3,41}$,
E.~Alpatov$^{41}$,
A.~K.~Alshammri$^{33}$
% ...
}

\address{\rm{(STAR Collaboration)}}

% ...
\address{$^{3}$Alikhanov Institute for Theoretical and Experimental Physics NRC "Kurchatov Institute", Moscow 117218}
% ...
\address{$^{17}$Czech Technical University in Prague, FNSPE, Prague 115 19, Czech Republic}
% ...
\address{$^{32}$Joint Institute for Nuclear Research, Dubna 141 980}
\address{$^{33}$Kent State University, Kent, Ohio 44242}
% ...
\address{$^{41}$National Research Nuclear University MEPhI, Moscow 115409}
% ...
\address{$^{45}$Panjab University, Chandigarh 160014, India}
% ...
\address{$^{59}$Texas A\&M University, College Station, Texas 77843}
% ...
\address{$^{66}$Variable Energy Cyclotron Centre, Kolkata 700064, India}
% ...


\begin{abstract}
The correlation between the mean transverse momentum, $[p_{\mathrm{T}}]$, and the squared anisotropic flow, $v^{2}_{n}$, on an event-by-event basis has been suggested to be influenced by the initial conditions in heavy-ion collisions. We present measurements of the variances and covariance of $[p_{\mathrm{T}}]$ and $v^{2}_{n}$, along with their dimensionless ratio, for Au+Au collisions at various beam energies: $\sqrt{\textit{s}_{NN}}$ $=$ 14.6, 19.6, 27, 54.4, and 200~GeV. Our measurements reveal a distinct energy-dependent behavior in the variances and covariances. In addition, the dimensionless ratio displays a similar behavior across different beam energies. We compare our measurements with hydrodynamic models and similar measurements from Pb+Pb collisions at the Large Hadron Collider (LHC). These findings provide valuable insights into the beam energy dependence of the specific shear viscosity ($\eta/s$) and initial-state effects, allowing for differentiating between different initial-state models.
\end{abstract}
\end{frontmatter}
\begin{multicols}{2}
Extensive experimental measurements of heavy-ion collisions at the Relativistic Heavy Ion Collider (RHIC) ~\cite{STAR:2005gfr, PHENIX:2004vcz, BRAHMS:2004adc, PHOBOS:2004zne} and the Large Hadron Collider~\cite{Muller:2012zq} demonstrate the formation of the Quark-Gluon Plasma (QGP), as predicted by Quantum Chromodynamics (QCD). Many of these investigations seek to characterize the QGP's transport properties (e.g.,  the specific shear viscosity $\eta / \textit{s}$)~\cite{Shuryak:1978ij, Shuryak:1980tp}.
The anisotropic flow measurements, which describe the azimuthal anisotropy of the particles
emitted relative to collision symmetry planes, are important in many studies~\cite{Shuryak:2003xe, Luzum:2008cw, Bozek:2009dw, Danielewicz:1998vz, Heinz:2001xi, Hirano:2005xf}.  Specifically,  anisotropic flow measurements reflect the viscous hydrodynamic response to the initial spatial distribution formed in the collision at early stages~\cite{Huovinen:2001cy, Hirano:2002ds, Romatschke:2007mq, Luzum:2011mm, Song:2010mg, Qian:2016fpi, Schenke:2011tv, Magdy:2020gxf, Magdy:2022ize}.

The anisotropic flow can be characterized by the Fourier expansion~\cite{Voloshin:1994mz,Poskanzer:1998yz} of the particle azimuthal angle ($\phi$) distributions,
\begin{eqnarray}
\label{eq:1-1}
\frac{dN}{d\phi} \propto   1+ 2 \sum^{\infty}_{n=1}\textit{v}_{n} \cos\left[  n (\phi - \Psi_{n})   \right]   ,
\end{eqnarray}
where $\Psi_{n}$ is the \nth-order flow symmetry plane.  The \nth complex anisotropic flow vector with \vn magnitude and $\Psi_{n}$ direction is defined as $\textit{V}_{n}$ $=$ $\textit{v}_n$ $e^{i \textit{n} \Psi_{n}}$. 
The first three flow coefficients are commonly called the directed flow(\first),  the elliptic flow(\second), and the triangular flow(\third).
Anisotropic flow measurements of the \nth-order flow harmonics~\cite{PHOBOS:2006dbo, STAR:2002hbo, STAR:2003xyj, PHENIX:2018wex, ALICE:2011ab,  ATLAS:2018ezv, CMS:2012tqw, CMS:2012zex, Magdy:2024ooh, Magdy:2019ojv, Adam:2019woz, Magdy:2018itt, Magdy:2023owx, Adamczyk:2017ird, Magdy:2017kji}, the correlation between different flow harmonics~\cite{STAR:2018fpo, ALICE:2016kpq, Aad:2015lwa, Qiu:2011iv, Adare:2011tg, Aad:2014fla, Aad:2015lwa}, and flow fluctuations~\cite{Magdy:2018itt, Alver:2008zza, Alver:2010rt, Ollitrault:2009ie} have led to a deeper understanding of the initial conditions of heavy-ion
collisions~\cite{Retinskaya:2013gca, Moreland:2014oya} and the properties of the matter created in heavy-ion collisions~\cite{Teaney:2003kp, Song:2011qa}.

Extensive comparisons between data and models across various collision energies and systems have indicated a low $\eta/s$ for the QGP, yet significant uncertainties persist~\cite{Shuryak:2003xe, Romatschke:2007mq, Luzum:2008cw, Bozek:2009dw, Song:2010mg, Shen:2011eg}. These uncertainties primarily arise from the difficulty in constraining the initial conditions of the collision. As a result, current theoretical and experimental investigations have sought to identify observables that could offer additional insights and tighter constraints on the initial state of heavy-ion collisions. One recent proposal is to examine the correlation between an event mean transverse momentum, $[p_{\mathrm{T}}]$, and the squared anisotropic flow, $v^{2}_{n}$, on an event-by-event basis~\cite{Giacalone:2020byk, Bozek:2016yoj, Schenke:2020uqq, Giacalone:2020dln, Lim:2021auv, ATLAS:2021kty}. These correlations are quantified using the Pearson correlation coefficient;
\begin{eqnarray}
    \rho(v^{2}_{n},[p_{T}]) = \frac{{\rm Cov}(v_{n}^{2},[p_T])}{\sqrt{{\rm Var}(v_{n}^{2})} \sqrt{{\rm Var}([p_T])}}\label{eq:1},
\end{eqnarray}
where,  $\rm{Cov(X,Y)}$ and $\rm{Var(X)}$ represent the covariance of parameters $X$ and $Y$ and the variance of parameter $X$ respectively.

The correlations between the $[p_T]$ and the $v^{2}_{n}$ coefficients offer valuable insights into two key aspects: (i) the relationship between the size and eccentricities in the initial state, and (ii) the correlation between the strength of the hydrodynamic response and the flow coefficients. It is noteworthy that $v^{2}_{n}$ ($\rm{Var(v^{2}_{n})}$) is primarily influenced by eccentricities, while $[p_T]$($\rm{Var([p_T])}$) is driven by the transverse size of the collision zone. Specifically, events with similar energy density but smaller transverse size tend to exhibit more radial expansion, leading to larger values of $[p_T]$($\rm{Var([p_T])}$)~\cite{Bozek:2012fw}. Furthermore, several studies~\cite{Giacalone:2019pca, Giacalone:2020awm, Giacalone:2021udy, Magdy:2022cvt, ATLAS:2021kty, STAR:2024eky, Zhang:2025hvi, STAR:2025elk} have highlighted the sensitivity of the ${\rho(v^{2}_{n},[p_{T}])}$ correlation to the initial-state nuclear deformation and the nuclear geometry of the colliding nuclei.

In this letter, we use the multiparticle cumulant method to study $\rho(v^{2}_{2},[p_{T}])$ and its components (${{Var}(v_{n}^{2})}, {Var}([p_T])$ and ${Cov}(v_{n}^{2},[p_T])$), dependence on beam energy. These measurements are expected to provide a unique data set that is comprehensive and derived from a consistent analysis across the different beam energies presented.

The Au+Au data analyzed in this study were collected by the STAR detector~\cite{Beddo:1992pv} using a minimum-bias trigger at center-of-mass energies \(\sqrt{s_{NN}} = 14.6\), 19.6, 27.0, 54.4, and 200~GeV, during the years 2019, 2019, 2018, 2017, and 2011, respectively. Collision vertices were reconstructed using charged-particle tracks measured by the STAR Time Projection Chamber (TPC), which operated in a 0.5~T magnetic field aligned with the beam (z) axis~\cite{Anderson:2003ur}. Events were required to have a longitudinal vertex position, \(V_z\), within \(\pm 30\)~cm for \(\sqrt{s_{NN}} = 200\)~GeV, \(\pm 40\)~cm for 54.4 and 27~GeV, and \(\pm 70\)~cm for 19.6 and 14.6~GeV. The transverse vertex position was constrained to \(r = \sqrt{V_x^2 + V_y^2} < 2\)~cm to ensure proximity to the beam axis. These vertex selection criteria were optimized for each energy to balance statistical needs with uniform detector performance: looser cuts at lower energies help maximize event yield, while tighter cuts at higher energies ensure consistent acceptance. Although such variations may influence the pseudorapidity coverage, their impact is mitigated by applying corrections as functions of \(V_z\), centrality, \(\eta\), and \(p_T\), and by incorporating any residual effects into the systematic uncertainty estimates.

In this work, the data sets are categorized into centrality classes based on the ``reference multiplicity" in the STAR experiment. This reference multiplicity is typically calculated from the raw multiplicity of primary charged particles reconstructed in the TPC across the entire azimuthal range and within $|\eta|$ $<$ 0.5. The selection of $|\eta|$ $<$ 0.5 is deliberate as it ensures a nearly uniform distribution of $\eta$ in this range of the detector, and the tracks exhibit higher quality compared to those near the TPC edge. Centrality classes are established by fitting the reference multiplicity distribution to that derived from Monte Carlo Glauber simulations~\cite{Adamczyk:2012ku, Abelev:2009bw}.

The $\rho(v^{2}_{n},[p_{T}])$ correlator is derived from the covariance and variance (cf. Eq.~\ref{eq:1}) which involve both two- and multi-particle correlations. The two- and multi-particle correlations can be influenced by nonflow effects due to resonance decays, Bose-Einstein correlations, and the fragments of individual jets~\cite{Jia:2013tja}. The nonflow contributions can be categorized into long-range nonflow, such as the awayside jet, and short-range nonflow arising from particles emitted within a localized region in pseudorapidity, $\mathrm{\eta}$. The latter can be effectively suppressed via sub-event cumulant methods~\cite{Jia:2017hbm, Huo:2017nms, Zhang:2018lls, Magdy:2020bhd, Zhang:2021phk}. In the sub-event cumulant methods, the correlated particles are taken from two or more sub-events that are separated in $\mathrm{\eta}$. The efficiency of these methods to reduce nonflow effects has been quantified for many different two- and multi-particle correlators~\cite{Jia:2017hbm, Huo:2017nms, Magdy:2020bhd, Zhang:2021phk, Magdy:2022ize, Magdy:2022jai}.

We construct both traditional (single-subevent, no $\eta$ separation) and subevent cumulants using charged particles with $0.2<p_{T}<2.0$~GeV. To correct for detector acceptance and tracking efficiency, we apply per-particle weights following Refs.~\cite{Jia:2017hbm, Magdy:2022ize, Magdy:2024eci}. These weights depend on centrality, $V_z$, $p_{T}$, $\eta$, and $\phi$, and incorporate the single-track efficiency.
The $\eta$ range of $0.35<|\eta|<1.0$ is selected to measure the two-particle $v_{n}\{2\}$ and the four-particle $v_{2}\{4\}$ using the two-subevent and traditional cumulant methods, respectively. Two subevents, A and C, are defined, where all tracks in sub-event A are required to be within $-1.0< \eta_{A}<-0.35$ and all tracks in sub-event C are required to be within $0.35<\eta_{C}<1.0$. The $v_{n}^{2}$ dynamical variance is given as~\cite{Bozek:2021zim}:
\begin{eqnarray}\label{eq:2-1}
    {\rm Var}(v_{n}^{2})_{dyn} &=& v_{n}\{2\}^{4} - v_{n}\{4\}^{4}  \nonumber,\\
                         &=& c^{n}_{2}\{2\} - c_{n}\{4\}.
\end{eqnarray}
It is important to note that, for the case of $n=3$, the four-particle cumulant $c_{3}\{4\}$ is found to be approximately zero at RHIC~\cite{STAR:2013qio}. The two-particle flow harmonics $v_{n}\{2\}$ can be given as:
\begin{eqnarray}\label{eq:2-2}
c_{n}\{2\} &=&  \langle \langle 2_n\rangle\rangle|_{AC}   =  \langle  \langle e^{\textit{i}~ n (\varphi^{A} -  \varphi^{C} )} \rangle \rangle, \nonumber \\ 
v_{n}\{2\}     &=&  \sqrt{c_{n}\{2\}},
\end{eqnarray}
where $\varphi_{A(C)}$ is the azimuthal angle of particles in the region $A$($C$) and {\color{black} $\langle \langle 2_n\rangle\rangle$ is the two particle correlations of order $n$}. In addition, the four-particle flow harmonic $v_{2}\{4\}$ can be given as:
\begin{eqnarray}\label{eq:2-3}
c_{2}\{4\}    &=&  \langle \langle 4_2\rangle\rangle - 2 \langle \langle 2_2\rangle\rangle^{2}, \nonumber \\
              &=& \langle  \langle e^{\textit{i}~ 2 (\varphi_{1} + \varphi_{2} -  \varphi_{3} -  \varphi_{4})} \rangle\rangle \nonumber  \\
              &-& ~2~ \langle  \langle e^{\textit{i}~ 2 (\varphi_{1} - \varphi_{2} )} \rangle\rangle \nonumber \\
v_{2}\{4\}    &=& \sqrt[4]{-c_{2}\{4\}},
\end{eqnarray}
where $\langle \langle 4_2\rangle\rangle$ is the four particle correlations of order $2$.

The dynamical $p_{T}$ correlations ($C_k$)~\cite{Bozek:2016yoj,Abelev:2014ckr,ALICE:2014gvd} in region $B$ ($|\eta_{B}|$ $<$ $0.35$) are defined as:
\begin{eqnarray}\label{eq:2-5}
\begin{aligned}
    C_{k} ={} & \langle  \frac{1}{N_{\rm pair}} \sum_{b}\sum_{b^{\prime}\neq b}  \\
              &   (p_{T,b} - \langle [p_T] \rangle )  (p_{T,b^{\prime}} - \langle [p_T] \rangle) \rangle,
\end{aligned}
\end{eqnarray}
where $\langle \rangle$ denotes an average over all events, and the two particles $b$ and $b^{\prime}$ are required to be separated by an $\eta$ gap $|\Delta\eta|>0.2$ to suppress short-range correlations, particularly from particle/resonance decays, that can contribute to $[p_T]$ and $C_k$. The coefficients \( C_k \), as defined in Eq.~\ref{eq:2-5}, encapsulate the event-by-event nature of (\( p_T \)) fluctuations. The event mean $p_T$, $[p_T]$,  is given as,
\begin{eqnarray}\label{eq:2-6}
     [p_T]  =  \sum^{M_{B}}_{i=1} p_{T,i}  /  M_{B},
\end{eqnarray}
where $M_{B}$ is the event multiplicity in  sub-event $B$. It is important to note that $[p_T]$ is measured within the same $\eta$ range used to define centrality. The potential impact of this overlap on the measured $[p_T]$ will be systematically explored in future studies.
The covariance between $v_{2}^{2}$ and $[p_T]$, ${\rm Cov}(v_{2}^{2},[p_T])$,  is defined using the three-subevents method~\cite{Aad:2019fgl,Zhang:2021phk} as,
\begin{eqnarray}\label{eq:2-7}
{\rm Cov}(v_{n}^{2},[p_T]) = \left\langle  \langle 2_n \rangle|_{AC} \left( [p_T] - \langle [p_T] \rangle \right)_{B} \right\rangle. 
\end{eqnarray}
The  resulting $\rho(v^{2}_{2},[p_{T}])$ correlator~\cite{Giacalone:2020byk,Lim:2021auv,Bozek:2016yoj,Bozek:2020drh,Schenke:2020uqq,Giacalone:2020dln,ATLAS:2021kty}, is obtained via Eqs.~\ref{eq:2-1}, \ref{eq:2-5} and \ref{eq:2-7}:
\begin{eqnarray}\label{eq:2-8}
    \rho(v^{2}_{n},[p_{T}]) = \frac{{\rm Cov}(v_{n}^{2},[p_T])}{\sqrt{{\rm Var}(v_{n}^{2})_{dyn} ~ C_k} }.
\end{eqnarray}
%

The presented measurement's systematic uncertainties are evaluated from variations in the analysis cuts for event selection, track selection, and nonflow suppression. (i) The systematic errors due to the event selection were obtained by modifying the vertex selection criteria to require the vertex position to be in either the positive or negative half of the allowed region along the beam. (ii) Track selection was varied by (a) reducing the distance of the closest approach (DCA) between a track and the primary vertex from its nominal value of 3~cm to 2~cm and (b) increasing the number of TPC space points used from more than $15$ points to more than $20$ points (out of a maximum of 45 space points per track in the STAR TPC). (iii) The pseudorapidity gap, $\Delta\eta~=~\eta_{1}-\eta_{2}$ for the track pairs, used to reduce the nonflow effects due to resonance decays, Bose-Einstein correlations, and the fragments of individual jets, was varied from $\Delta\eta = 0.7$ to  $\Delta\eta = 0.9$. The overall systematic uncertainty, assuming independent sources, was taken to be the quadrature sum of the uncertainties resulting from the respective cut variations. The systematic uncertainties depend on both beam energy and collision centrality, and vary from 5\% to 9\%. The $\Delta\eta$ variation is the dominant source of systematic uncertainty.

The comprehensive beam energy measurements reported in this paper will help theoretical models constrain their initial conditions and better estimate how \etas varies with baryon chemical potential and temperature.
Hence, it is instructive to compare our measurements to available theoretical predictions. In this work, our measurements at \roots = 200 GeV are compared to two theoretical models: (i) The IP-Glasma + (3+1)D MUSIC + UrQMD model~\cite{Schenke:2019ruo, Schenke:2020uqq}, which utilizes an initial state generated from the IP-Glasma framework, followed by the MUSIC viscous hydrodynamic model with \etas of 0.12, and the UrQMD afterburner. (ii) The Trento+Hydro model~\cite{Alba:2017hhe}, which employs initial conditions from the T$_R$ENTo model with a viscous hydrodynamic model with \etas of 0.05. In both models, the \etas value selection was chosen to reproduce the experimental measurements of the anisotropic flow and particle spectra.

 \begin{figure}[H]
  \centering{
 \includegraphics[width=1.0 \linewidth, angle=0]{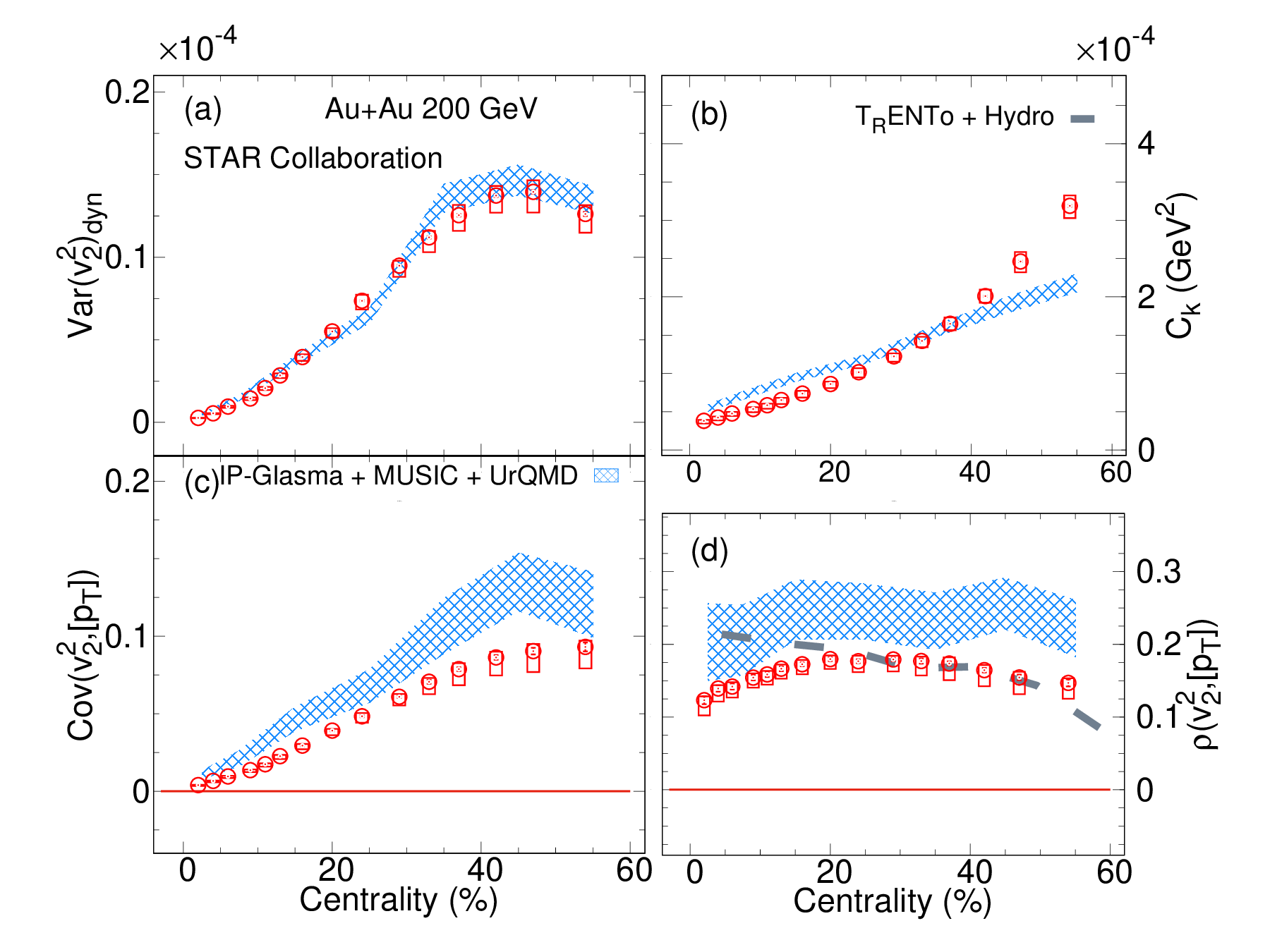}
\caption{
Comparison of the centrality dependence of (a) ${\rm Var}(v_{2}^{2})_{dyn}$, (b) $C_{k}$,  (c) ${\rm Cov}(v_{2}^{2},[p_T])$, and  (d) $\rho(v^{2}_{2},[p_{T}])$, for Au+Au collisions at \roots = 200~GeV.  Measurements are compared to theoretical calculations represented as hatched bands~\cite{Schenke:2019ruo} and dashed curve~\cite{Alba:2017hhe}.  \label{Fig:1}
} }
 \end{figure}
Figure~\ref{Fig:1} shows the centrality dependence of (a) ${\rm Var}(v_{2}^{2})_{\rm dyn}$, (b) $C_{k}$, (c) ${\rm Cov}(v_{2}^{2},[p_T])$, and (d) $\rho(v_{2}^{2},[p_{T}])$ for Au+Au at \roots=200~GeV. The shaded bands in panels (a)–(d) denote viscous hydrodynamic predictions from IP-Glasma + MUSIC + UrQMD ~\cite{Schenke:2019ruo}, and the dashed curve in panel (d) shows a hydrodynamic calculation with T$_R$ENTo initial conditions~\cite{Alba:2017hhe}. The IP-Glasma+MUSIC+UrQMD calculation to a good degree describes ${\rm Var}(v_{2}^{2})_{\rm dyn}$ and $C_{k}$, but overpredicts ${\rm Cov}(v_{2}^{2},[p_T])$ and $\rho(v_{2}^{2},[p_{T}])$. For $\rho(v_{2}^{2},[p_{T}])$, the T$_R$ENTo calculation agrees with the data for centralities $\gtrsim 20\%$, but overpredicts in the 0–20\% most central collisions, where it is consistent with the MUSIC-based prediction. 
The differing magnitudes and trends underscore the constraining power of $\rho(v_{2}^{2},[p_{T}])$ on theoretical models and motivate more detailed data--model comparisons in future work.

 \begin{figure}[H]
  \centering{
\includegraphics[width=1.0 \linewidth,angle=0]{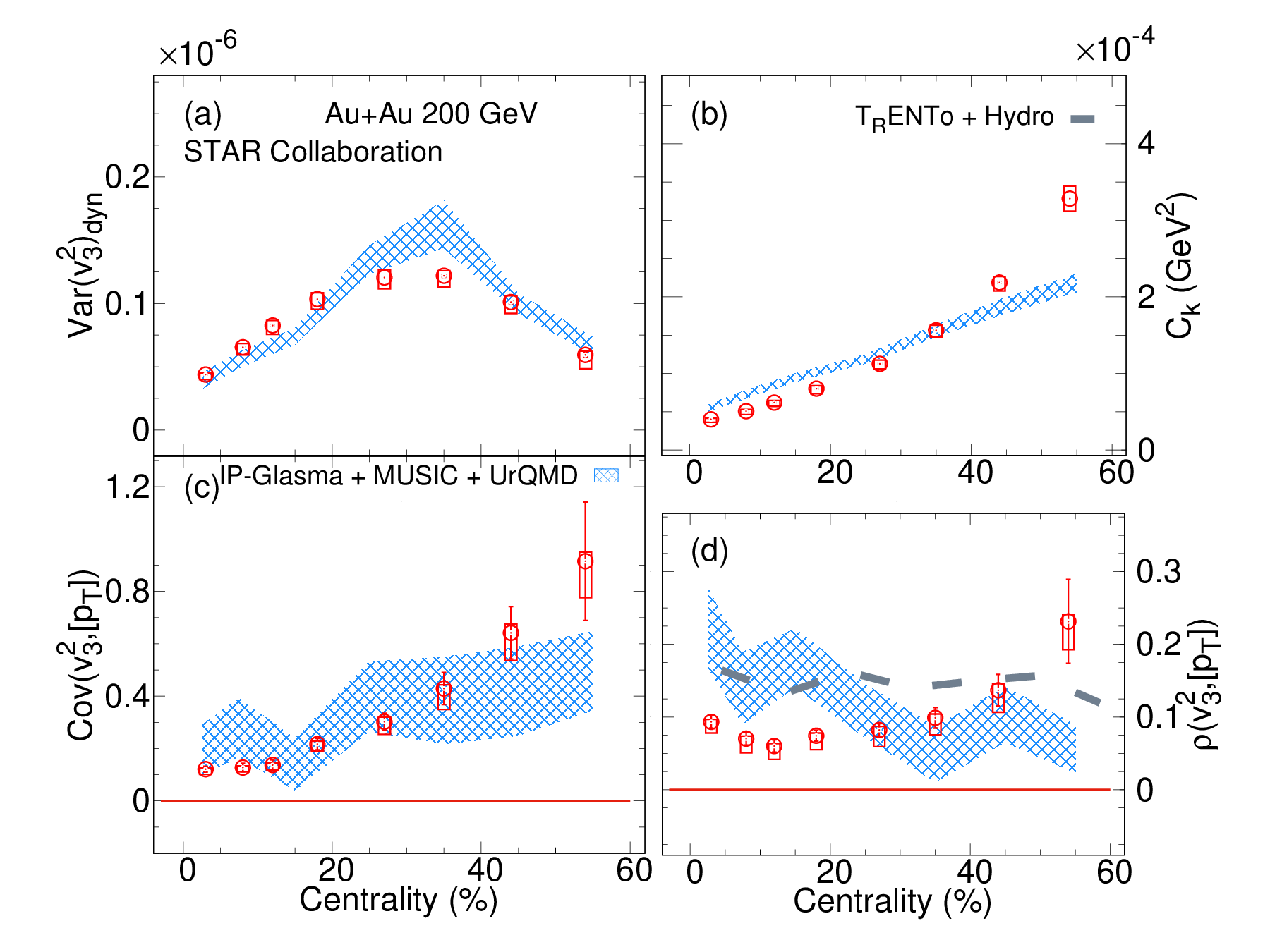}
\caption{
Comparison of the centrality dependence of (a) ${\rm Var}(v_{3}^{2})_{dyn}$, (b) $C_{k}$,  (c) ${\rm Cov}(v_{3}^{2},[p_T])$, and  (d) $\rho(v^{2}_{3},[p_{T}])$, for Au+Au collisions at \roots = 200~GeV.  Measurements are compared to theoretical calculations represented as hatched bands~\cite{Schenke:2019ruo} and dashed curve~\cite{Alba:2017hhe}.\label{Fig:2}
 }
}
\end{figure}
Figure~\ref{Fig:2} shows the centrality dependence of the  ${\rm Var}(v_{3}^{2})_{dyn}$ (a), $C_{k}$ (b), ${\rm Cov}(v_{3}^{2},[p_T])$ (c), and $\rho(v^{2}_{3},[p_{T}])$ (d) for Au+Au at \roots = 200 GeV. The shaded bands in panels (a)--(d) and the dashed curve in panel (d) show the viscous hydrodynamic calculations of the presented quantities~\cite{Alba:2017hhe, Schenke:2019ruo}. The theoretical calculations~\cite{Schenke:2019ruo} show reasonable agreement with the ${\rm Var}(v_{3}^{2})_{dyn}$, $C_{k}$, and the ${\rm Cov}(v_{3}^{2},[p_T])$, however, they over predicts the $\rho(v^{2}_{3},[p_{T}])$ measurements in central collisions.

The comparison between data and model calculations in Figs.~\ref{Fig:1} and \ref{Fig:2} shows that, while current models provide a reasonable description of ${\rm Var}(v_{n}^{2}){\rm dyn}$, they tend to overestimate the values of $\rho(v{n}^{2},[p_T])$ in central collisions, especially for $n = 3$. This discrepancy suggests that the existing models may be missing key dynamical ingredients or correlations that become increasingly important in the most central events. These results highlight the need for further theoretical developments to improve the modeling of the interplay between flow fluctuations and transverse momentum dynamics. More broadly, they underscore the power of these measurements to provide stringent constraints on theoretical frameworks, extending their impact beyond the conventional description of $v_n$.

Studying the beam energy dependence of the observables presented in this work offers a valuable opportunity to disentangle medium properties from initial-state effects in heavy-ion collisions~\cite{Magdy:2021cci, Giacalone:2020awm, Magdy:2021ocp, Schenke:2020uqq, Zhang:2025yyd}. The analysis incorporates both single-component observables, ${\rm Var}(v_{n}^{2}){\rm dyn}$, ${\rm Var}([p_T])$, and ${\rm Cov}(v{3}^{2},[p_T])$, as well as dimensionless correlation measures such as $\rho(v_{n}^{2},[p_T])$. Previous investigations have shown that single-variable observables are sensitive to the interplay between initial- and final-state effects, which are expected to evolve with beam energy due to changes in the medium's temperature, density, and lifetime. In contrast, dimensionless observables like $\rho(v_{n}^{2},[p_T])$ have been proposed as more direct probes of the initial-state geometry and its event-by-event fluctuations. Model calculations indicate that these correlation measures are largely insensitive to $\eta/s$ and are instead dominated by initial-state effects~\cite{Magdy:2021cci, Virta:2024avu, Giacalone:2020byk}. This separation of sensitivities provides a powerful, testable hypothesis: since prior measurements at RHIC suggest that initial-state effects change only weakly with beam energy~\cite{STAR:2022gki}, dimensionless observables should show little to no dependence on collision energy. In contrast, observables sensitive to $\eta/s$ are expected to exhibit a stronger dependence on beam energy, reflecting changes in the system's dynamical evolution. By measuring both classes of observables across a range of beam energies, this study aims to test these expectations and to provide new insights into the interplay between initial-state effects and medium transport properties. Ultimately, this combined analysis serves to validate the assumed sensitivity hierarchies and supports a more comprehensive understanding of the system created in heavy-ion collisions.

 \begin{figure}[H]
   \centering{
\includegraphics[width=1.0 \linewidth,angle=0]{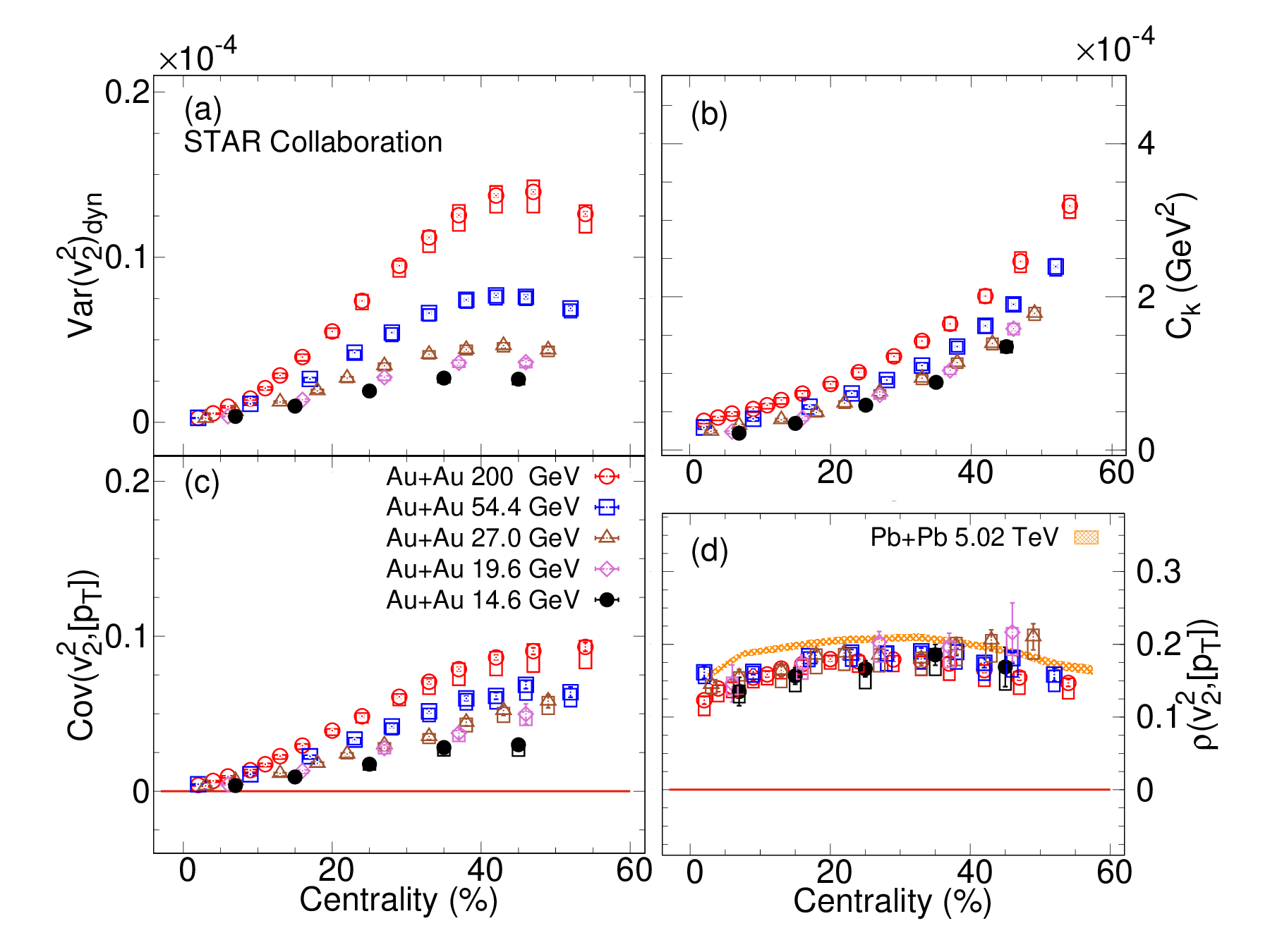}
\caption{
Comparison of the centrality dependence of (a) ${\rm Var}(v_{2}^{2})_{dyn}$, (b) $C_{k}$,  (c) ${\rm Cov}(v_{2}^{2},[p_T])$, and  (d) $\rho(v^{2}_{2},[p_{T}])$, for Au+Au collisions at \roots = 200, 54.4, 27, 19.6, and 14.6~GeV. The curve represents the LHC measurements~\cite{ALICE:2021gxt}. \label{Fig:3}
} 
}
\end{figure}
Figure ~\ref{Fig:3} shows the centrality dependence of ${\rm Var}(v_{2}^{2})_{dyn}$ (a), $C_{k}$ (b), ${\rm Cov}(v_{2}^{2},[p_T])$ (c), and $\rho(v^{2}_{2},[p_{T}])$  (d) at five different beam energies. The measurements in panels (a)--(c) show that the magnitudes of ${\rm Var}(v_{2}^{2})_{dyn}$, $C_{k}$, and ${\rm Cov}(v_{2}^{2},[p_T])$ decreases with beam energy over the full centrality range. Such behavior would reflect the beam energy dependence of the final state effect, given that the initial state is not a strong function of beam energy~\cite{Magdy:2021ocp, Magdy:2021cci, STAR:2022gki, STAR:2022vkx}. In contrast,  the values for  $\rho(v^{2}_{2},[p_{T}])$ panel (d) show weak beam energy dependence. The $\rho(v^{2}{2},[p{T}])$ measurements are compared to published LHC results at $\sqrt{s_{\mathrm{NN}}} = 5.02$ TeV~\cite{ALICE:2021gxt}. A small difference is observed between the two measurements, which may be caused by variations in centrality definitions, mean transverse momentum, and track selection criteria (e.g., $p_T$ and $\eta$ cuts) between the two experiments. A detailed, matched-condition comparison is beyond the scope of this work, which focuses on RHIC energies with selections optimized for statistical precision in the RHIC data. Nonetheless, the overall consistency supports a largely energy-independent behavior of $\rho(v^{2}{2},[p{T}])$ within current uncertainties.

 \begin{figure}[H]
 \centering{
 \includegraphics[width=1.0 \linewidth, angle=0]{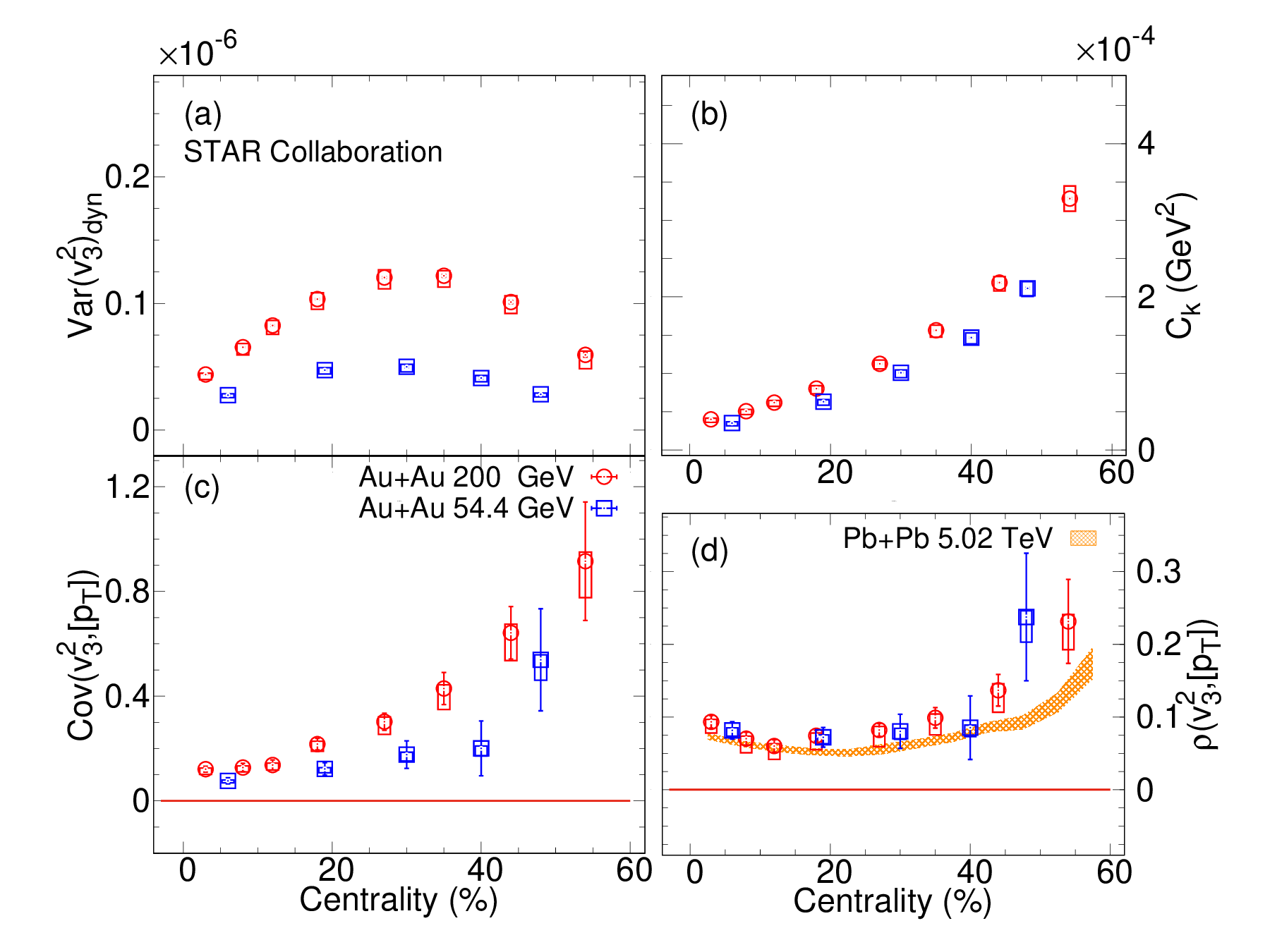}
 \caption{
 The centrality dependence of the ${\rm Var}(v_{3}^{2})_{dyn}$ (a), $C_{k}$ (b), ${\rm Cov}(v_{3}^{2},[p_T])$ (c), and $\rho(v^{2}_{3},[p_{T}])$ (d), for Au+Au collisions at \roots = 200 and 54.4~GeV. The curve represents the LHC measurements~\cite{ALICE:2021gxt}. \label{Fig:4}
 }
}
 \end{figure}
The centrality and beam-energy dependence of ${\rm Var}(v_{3}^{2})_{dyn}$ (a), $C_{k}$ (b), ${\rm Cov}(v_{3}^{2},[p_T])$ (c), and $\rho(v^{2}_{3},[p_{T}])$ (d) are presented in Fig.~\ref{Fig:4}. Our presented measurements in Fig.~\ref{Fig:4} are shown at high energies (i.e., \roots = 200 and 54.4~GeV) due to statistical limitations at lower energies. The $\rho(v^{2}_{3},[p_{T}])$ panel (d) are compared to similar LHC measurements~\cite{ALICE:2021gxt}. Similar to the measurements in Fig.~\ref{Fig:3}, the magnitudes of ${\rm Var}(v_{3}^{2})_{dyn}$ and ${\rm Cov}(v_{3}^{2},[p_T])$ decrease with beam energy. In contrast, the $\rho(v^{2}_{3},[p_{T}])$ panel (d) shows weak, if any, beam energy dependence. The $\rho(v^{2}{3},[p{T}])$ measurements are also compared to corresponding LHC results~\cite{ALICE:2021gxt}, showing good agreement and reinforcing the consistency of this observable across collision energies.

In summary, we present a comprehensive analysis of event-by-event correlations between mean transverse momentum and squared anisotropic flow coefficients in Au+Au collisions across a broad range of beam energies ($\sqrt{s_{NN}} = 14.6$--$200$~GeV) using the STAR detector at RHIC. Our study examines both single-component observables, such as ${\rm Var}(v_n^2)_{\rm dyn}$, ${\rm Var}([p_T])$, and ${\rm Cov}(v_n^2, [p_T])$, and the dimensionless correlation measure $\rho(v_n^2, [p_T])$ for $n = 2, 3$, motivated by theoretical expectations of their different sensitivities to initial- and final-state effects. The results show a clear beam energy dependence for the single-component observables, consistent with changes in the medium's dynamics, while $\rho(v_n^2, [p_T])$ remains nearly independent of energy, reflecting its sensitivity to initial-state geometry. Comparisons to leading theoretical models reveal reasonable agreement for the single-component observables but a consistent overprediction of $\rho(v_n^2, [p_T])$, especially in central collisions, indicating that current models may be missing key dynamical features. Collectively, these findings validate the proposed separation in sensitivity between initial- and final-state effects, provide new constraints for theoretical modeling, and highlight the value of combined observables in advancing our understanding of the quark-gluon plasma created in heavy-ion collisions at RHIC.

We thank the RHIC Operations Group and SDCC at BNL, the NERSC Center at LBNL, and the Open Science Grid consortium for providing resources and support.  This work was supported in part by the Office of Nuclear Physics within the U.S. DOE Office of Science, the U.S. National Science Foundation, National Natural Science Foundation of China, Chinese Academy of Science, the Ministry of Science and Technology of China and the Chinese Ministry of Education, NSTC Taipei, the National Research Foundation of Korea, Czech Science Foundation and Ministry of Education, Youth and Sports of the Czech Republic, Hungarian National Research, Development and Innovation Office, New National Excellency Programme of the Hungarian Ministry of Human Capacities, Department of Atomic Energy and Department of Science and Technology of the Government of India, the National Science Centre and WUT ID-UB of Poland, German Bundesministerium f\"ur Bildung, Wissenschaft, Forschung and Technologie (BMBF), Helmholtz Association, Ministry of Education, Culture, Sports, Science, and Technology (MEXT), Japan Society for the Promotion of Science (JSPS), and Agencia Nacional de Investigacion y Desarrollo de Chile (ANID), Chile.

\bibliographystyle{elsarticle-num}
\bibliography{ref} 
\end{multicols}
\end{document}